\documentclass[preprint]{JHEP3} 
\usepackage{epsfig,multicol}

\newcommand\fverb{\setbox\pippobox=\hbox\bgroup\verb}
\newcommand\fverbdo{\egroup\medskip\noindent%
            \fbox{\unhbox\pippobox}\ }
\newcommand\fverbit{\egroup\item[\fbox{\unhbox\pippobox}]}
\newbox\pippobox                                                                  %
\def\){\right)}
\def\({\left( }
\def\]{\right] }
\def\[{\left[ }

\def\half{{1\over 2}}

\newcommand{\be}{\begin{equation}}
\newcommand{\ee}{\end{equation}}
\newcommand{\ba}{\begin{eqnarray}}
\newcommand{\ea}{\end{eqnarray}}

\newcommand{\ket}{\rangle}
\newcommand{\bra}{\langle}

\def\Tr{{\rm Tr}}

\title{Holography of Radiation and Jet Quenching}

\author{Sang-Jin Sin$^{a, b}$   and  Ismail Zahed $^c$\\
$^{a}$ Department of Physics, Hanyang University, 133-791, Seoul, Korea\\
$^b$ High Energy section, ICTP, Strada Costeria, 11-34014 Trieste,
Italy \\
$^c$ Department Physics and Astronomy, State university of New
York, Stony Brook, NY11794 }

\preprint{IC/IR/2004/6}

\abstract{We study the non-linear propagation of radiation in
{\cal N=4} SYM at zero and finite temperature using the refined
radius/scale duality in AdS/CFT. We argue that  a pulse radiation by a quark at the boundary should be described holographically by a ``point like object" passing through the center of the AdS bulk.  We find that at finite 
temperature, the radiation stalls at a distance of $1/\pi T$ with
a natural geometric and holographic interpretation. Indeed, the
stalling is the holographic analogue of the gravitational in-fall
of light towards the black hole in the bulk.   We suggest that these results are relevant for jet quenching by a strongly coupled quark-gluon liquid as currently probed in heavy ion colliders at RHIC. In particular, colored jets cannot make it beyond $1/3$ fm at RHIC whatever their energy.}


\begin{document}
\section{Introduction and summary}
One of the most important achievement in modern day's string
theory is the duality relation between ${\cal N}=4$ super
Yang-Mill (SYM) theory and the gravity theory in Anti-De Sitter
space\cite{adscft}. The calculations of the Wilson lines of 3+1
SYM in vacuum \cite{rey,malda} and at finite temperature
\cite{witten2,sonn,rey2} have led to insightful results for the
static potential between the quark and anti-quark at strong
coupling~\cite{sz1}. In particular a new form of Coulomb's law
operates at strong coupling. While the general results show that
there is no confinement in ${\cal N}=4$, one can still
construct various deformation of it having the area law of Wilson
lines with some supersymmetry broken. At finite temperature, one
still has a screening behavior but with no exponential tail, which is a remarkable feature of a strongly coupled gauge theory.
A further use of time-like Wilson lines was initiated in~\cite{rsz}
whereby small angle elastic parton-parton scattering was described
by a minimal surface with a boundary fixed by the world lines of
the parton-parton scatterers. The extension of this idea to the
inelastic case was described in~\cite{janik}.

Recently, it was suggested that QCD as probed by relativistic
heavy ion collisions at RHIC is in a deconfined but
strongly coupled phase~\cite{sz2}. The released quarks and gluons
are likely in a strongly coupled liquid after their prompt release 
rather than a weakly coupled quark-gluon plasma. The empirical
evidence for this stems from the necessity of early equilibration
on a time scale of 1 fm, to accommodate the observed and
exceedingly large flow of matter. Another important aspect of the
strongly coupled liquid released at RHIC is its large opacity to
high energy color probes or jet quenching. Weakly coupled
deconfined QCD cannot accommodate for the large jet quenching
reported at RHIC. Could jet quenching at RHIC be the result of jet
interaction in a strongly coupled quark-gluon liquid,
and if so how would we go about addressing it from first principles?

In this paper we would like to suggest a way to address the issue
of jet quenching in a strongly coupled QCD liquid as currently
probed by RHIC, by calculating its analogue in ${\cal N}=4$ SYM at
finite temperature and strong coupling. Since supersymmetry and
conformal symmetry are irrelevant at finite temperature, strongly
coupled ${\cal N}=4$ SYM and QCD just above the deconfinement
temperature share much in common~\cite{sz1,sz2}. Since jets at
RHIC are colored objects released in prompt parton-parton
collisions, zipping through a strongly interaction quark-gluon
liquid we need to formulate the problem of how colored waves get
depleted in hot ${\cal N}=4$ SYM. In other words, if one could argue that gluon radiation field can not propagate in the strong coupled hot medium, it should be enough to explain the depletion of Jets. Since the perturbative QCD calculation is out of reach for this problem and AdS/CFT is for strongly coupled Yang-Mill theory, it is natural to expect that the latter should give a fundamental understanding for the Jet quenching phenomena.

In order to address this problem, one need to know the nature of
the non-abelian wave-propagation in the hot medium as well as in
vacuum from the AdS/CFT point of view. While the static force in strong coupling SYM is well understood in terms of Wilson line probe, the radiation
problem is still largely open question. In fact, in ref.
\cite{callan} difficulties in getting radiation
field were emphasized. Namely, one point function $\Tr F^2$ following the AdS/CFT prescription did not give a piece representing the radiation field. 
More recently, a non-trivial progress in this 
direction was made  by Mikhailov~\cite{mik} using the
minimal surface constructed by ruled surface. He
showed that the power exchanged between $q\bar q$ pair
sitting at north and south poles of the boundary is exactly of the Lienard-type for the power radiated by a moving charge in
electrodynamics. After this success, it is a tempting question to ask the radiation for other situation where quark and antiquark are located in generic positions. We will argue that the minimal surface describing the radiation will be disjoint from the minimal surface that describe the static Coulomb force. In fact, the radiation emitted by a quark is absorbed by its image anti-quark at the antipodal point of the $S^3$ boundary rather than by a nearby anti-quark. In the flat boundary case, this means that radiation emitted by a quark
propagate to the infinity without being totally absorbed by other
anti-quark at finite distance, unlike the static flux.

We will study the non-abelian radiation using the holographic correspondence between light (gluon) propagation in bulk and boundary.  
For pure AdS back ground, we will derive a holographic correspondence between 'a point in the bulk' and 'a sphere in the boundary' by identifying the light propagation in the bulk and boundary.
In flat boundary case  this is  
reduced to the well known radius/size relation in ads/cft. 
For the black hole background light propagation in the bulk is still easy problem, while it is not a directly calculable problem in the boundary theory. 
In this situation, if we know the holographic correspondence between the bulk and boundary, calculating the boundary propagation would reduce to an another easy problem. 

To show above idea more concretely,  we propose a metric independent holographic correspondence. Using this point-sphere correspondence, we will show that the hot medium literally stops the propagation at a distance of $1/\pi T$. 
This stopping/screening of time-like
radiation is just the dual phenomenon of gravitational in-fall of
'light' towards the black hole in bulk. 
The stopping/screening length is the same whatever the energy carried by the wave 
including high energy colored jets. Strongly coupled hot ${\cal
N}=4$ SYM is totally opaque to color waves beyond a length scale
of $1/\pi T$. The arguments presented in~\cite{sz1,sz2} suggest
that this observation carries to QCD near its deconfinement
temperature as probed by RHIC. In other words, the strongly
coupled quark-gluon liquid triggered at RHIC is opaque to jets
whatever their energy. The opacity length is about $1/3$ fm at
RHIC.

The rest of the paper goes as follows. In section 2, we describe
how to describe the non-linear radiation in AdS/CFT. We notice  that the radiation at the boundary should be described in bulk by a point like object passing through the center of the bulk. In section 3, we study holography of radiation. The point in a bulk corresponds to the light front sphere in the boundary.  In section 4, we describe the geometry of screening length
by using the concepts prepared in sections 2 and 3. We conclude in section 5.

\section{ Radiation in the AdS/CFT }

The issue of radiation in AdS/CFT context was first  discussed quantitatively (but unsuccessfully) in \cite{callan}. The authors considered the field of charged oscillator by  
calculating  the  $\bra\Tr F_{\mu\nu}^2\ket $   using the Gubser-Klebanov-Polyakov-Witten(GKP-W) prescription and found that the result does not contain the effect of radiation fields since $\bra\Tr F_{\mu\nu}^2\ket \sim 1/r^4$ in large $r$ limit.
\footnote{In electromagnetism, the electric field of a moving particle has
two distinguishable parts: one is the near-field (electrostatic)
that falls as $1/r^2$ and depends  only on the velocity of the
source, the other is the far-field (radiation) that falls as $1/r$
and depends on the acceleration of the source. Far away, the
radiation part dominates.} 
Since it is unlikely that GKP-W prescription is invalid, they opened the possibility that there might be no radiation field in the strong coupled Yang-Mill system.  However one should notice  that even in the case there are radiation fields, we still get $\bra\Tr F_{\mu\nu}^2\ket \sim 1/r^4$. This is because $\bra\Tr F_{\mu\nu}^2\ket =\bra E^2\ket -\bra B^2\ket $ and $\bra E^2\ket =\bra B^2\ket \sim 1/r^2$ for radiation fields. Namely the leading radiation parts are cancelled while the static parts do not. The lesson we should learn is that since there is no easy way to calculate the electric and magnetic parts separately using the GKP-W prescription, we should lean on some other method to study the radiation. 

The first successful evidence for the existence of the radiation field came from the consideration of Wilson lines for the accelerating quark by Mikhailov~\cite{mik}.
Here we briefly summarize the work of Mikhailov. The set
up consists of a wiggly time-like Wilson line where the quark at
the North pole plays the role of a non-Abelian emitter and the
anti-quark at the South pole the receiver as shown in
Fig.~\ref{wiggle}. \DOUBLEFIGURE[t]{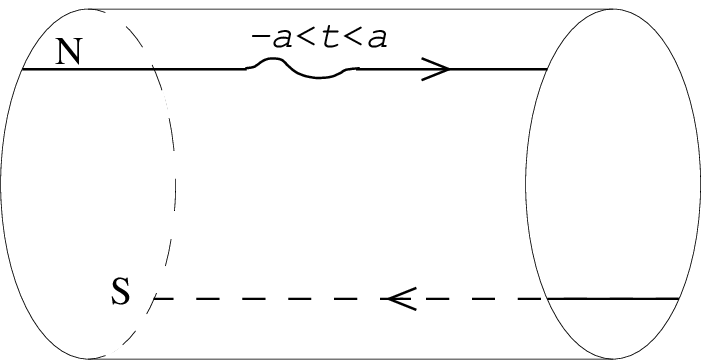,
width=.4\textwidth} {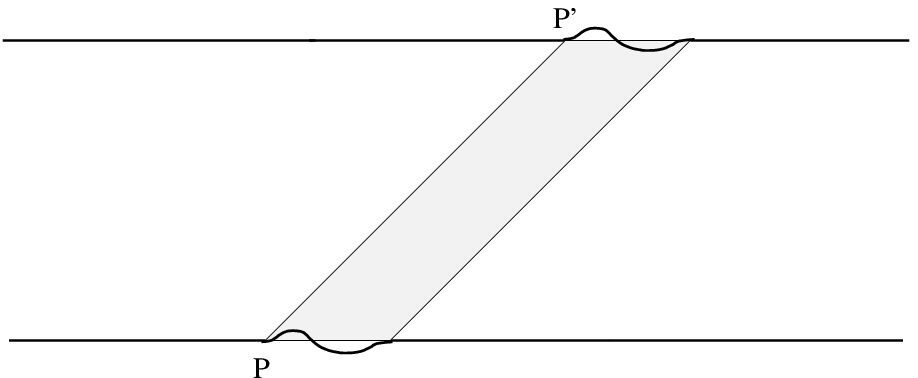, width=.4\textwidth}{Wiggly
time-like Wilson line where the quark at the North pole plays the
role of a non-Abelian emitter and the anti-quark at the South pole the role of a receiver. \label{wiggle}}{The minimal surface
describing the radiation contains the `light' strip. It has the
wiggly $q \bar q$ lines as its boundaries. \label{lightstrip}}
Throughout we will refer to the non-Abelian gluon propagation in
vacuum and at finite temperature as 'light' propagation to appeal
to intuition. The minimal surface describing the process has  a
strip. See Fig. \ref{lightstrip}. We call the wiggly strip in the
minimal surface as the light-strip. He observed that this minimal
surface can be described as a ruled surface in $\bf R^{2+4}$ in
which the $AdS_5$ is embedded as a hyperbolic sphere and the
boundary of the surface is fixed by the history of the source of
the radiation. Using this, he has shown that the non-Abelian power exchanged between these antennas is exactly of the Lienard-type:
 \be
E=\frac{\sqrt{\lambda}}{2\pi}\int dt
\frac{\ddot{\vec{x}}^2-[\dot{\vec{x}}\times
\ddot{\vec{x}}]^2}{\(1-\dot{\vec{x}}^2\)^3}, \label{energy} \ee
where $\lambda={4\pi g_s N}$ and the $x^i(t)$ is the trajectory of
the quark at the boundary.
 The only difference is the occurrence of $\sqrt{\lambda}/2\pi$ instead of
$2e^2/3$ as the overall factor. 
Although Mikhailov considered a special configuration where 
quark and an anti-quark are located at the North/South poles,
his light-like minimal surface captures the essentials of radiation. 

\subsection{Radiation for the generic  charge positions}
An urgent question to ask is how to describe the radiation  for 
generic two positions of a quark and an anti-quark at the
boundary. It is tempting to answer this question by considering the perturbation of minimal surface whose boundary is given by  
quark-antiquark Wilson lines with small wiggles. 
See {\bf right} figure of fig.\ref{light2}.
Although this might look plausible, we can argue that this can not be the right description with three evidences.  
\begin{enumerate} 
    \item  
The Mikhailov construction   for the light strip does not go through for this case. The reason for this is that the ruled surface is completely determined by fixing the quark line. There is no room to accommodate the anti-quark line as another boundary unless the the latter is located at the antipodal point of the former.
    
    \item     
Flight time analysis shows that the time for IL to arrive at the antiquark along the string at the bulk connecting the quark and the antiquark is different from the time for the light to propagate  from quark to antiquark. In order words, Lorentz invariance is broken if this is correct description for the light propagation. See the appendix of this section.

\item For this case  the emitted radiation is completely
        absorbed by the opposite charge. This seems impossible since corresponding statement in flat boundary is that  radiation is absorbed totally by a charge at the finite distance from the light source.  
\end{enumerate}

 What is the resolution for the problem? The third argument is a strong hint for the problem. The definition of the radiation field in the flat space is the piece of electric field  whose long distance behavior is $1/r$. This means that the intensity goes like $1/r^2$ so that the integral of the intensity over the sphere at infinite distance is finite. Therefore all the radiation flux should go to the point at infinity. In terms of global co-ordinate where the spacial part of the geometry is $S^3$, the addition of one point at infinity to $R^3$,  the radiation is absorbed by the 'image charge' at the antipodal point of the emitting charge. While
static field lines are totally absorbed by the antiquark nearby,
the latter does not play any role for the radiation and  therefore  the the minimal surface describing the radiation is disjoint from minimal surfaces that describe the static potential. See the left of figure of fig.\ref{light2}.
For the second problem listed above, it is also easy to show that  for the quark and antiquark located at the antipodal point of each other, the flight time in the bulk is identical to that in the boundary as shown in the appendix of this section. 

All these evidences support our proposal that for the radiation of a charge should be described by the charge-image charge at the antipodal points, regardless of charge configuration of the system.

\FIGURE{\epsfig{file=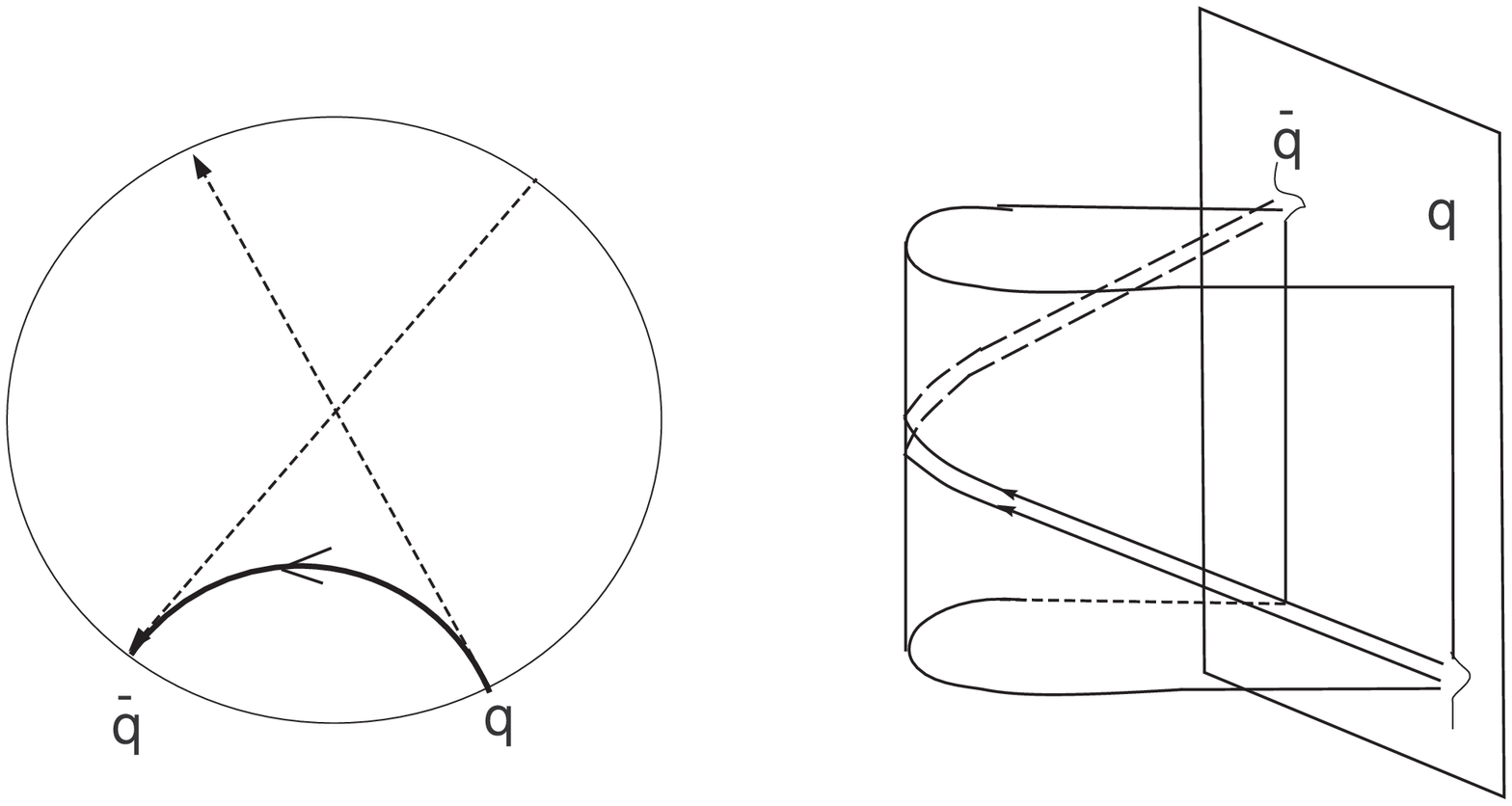,width=10cm}
        \caption[fig]{ 
        {\bf Left}:Radiation is absorbed by a image charge at 
      the antipodal point  regardless of nearby charge configuration.  Hence the light sheet (dotted lines) is
        disjoint from the static minimal surface (bold line).
          We draw trajectories rather than histories.
          {\bf Right}: Trial answer to the radiation problem for generic charge configuration, which turns out to be inconsistent.  \label{light2}}}
 
\subsection{Appendix to section 2: Flight time analysis}
In the scaling coordinate(s) ($r=R^2 U$ or $r=R^2/z$), the metric
and the action are given by \be ds^2=\alpha'R^2\[-dt^2+
U^2dx^2+{dU^2\over U^2}\] \label{r3bdym1}\ee The solution to the
minimal surface problem is given by \be x=U_0^2\int_{U_0}^U
\frac{dU}{U^2\sqrt{U^4-U^4_0}}, \;\;\; t=\int_{U_0}^U
\frac{dU}{\sqrt{U^4-U^4_0}}. \ee 
The separation between the emission and absorption in the boundary is given by  
\be
 \Delta x= {2\over U_0}\int_1^\infty\!\!\!
 \frac{dy}{y^2\sqrt{y^4-1}}={1\over 2U_0}B({3\over
 4},\half) .
 \ee
On the other hand the flight time $\Delta t$ in the bulk is given by solving the null geodesic. 
 \be
 \Delta t_{bdy}={2\over U_0}\int_1^\infty\!\!\!
 \frac{dy}{\sqrt{y^4-1}}= {1\over 2U_0}B({1\over 4},\half).
 \ee
Since two description (the bulk one and the boundary one) must coincide,  the velocity of the propagating gluon $v_g$  should be defined as 
the ratio of these two: 
\be
v={\Delta x\over\Delta t}= {8\pi^2\over \Gamma\!\!\({1\over4}\)^4
}\cong 0.457, \ee which is different from the light velocity 1.
Clearly this is in violation of Lorentz invariance, which is not
expected to be broken at strong coupling. This analysis is enough
to show that the second possibility mentioned above and shown in
Fig.~\ref{light2} is not likely to happen. Above calculation was
done in ref. \cite{rey3} to analyze propagation of the wave
created when one shakes the one end of the string along the
U-shaped string. The speed $0.457$ was interpreted in
\cite{callan} as subluminal mean speed. \footnote{Mikhailov  pointed out the ref.\cite{rey3} where   the flight time analysis in the flat boundary case is done.}

For the spherical boundary, one get a result where the travelling time that depends on the travel distance.  However,  when quark anti-quark are located at antipodal positions, we get \cite{suss-polch}
 \be
\Delta t=R\pi, \hskip0.5cm\Delta x=R\Delta \theta= R\pi,
\hskip.5cm {\rm and}\hskip0.5cm v=\frac{\Delta x}{\Delta t}=1. \ee
Namely, the boundary flight time is exactly equal to that in the bulk. This is another evidence that the wave
propagating along the string describe the radiation only when
their end points are antipodal points. Moreover we can say that
whether there is a nearby quark or not, radiation is described by
the wave propagating along the string connecting the emitting
quark and its image partner sitting at its antipodal point. This
is the picture indicated in the left of Fig. (\ref{light2}). Since
this corresponds to the flat space statement that radiation
propagate to the infinity, it makes perfect sense.

 \section{Holography of Radiation} 
The emerging picture of radiation in AdS/CFT is that in the bulk  it is described by a trajectory of a point like object passing through center of the AdS space, which we call 'idea of light', while in the boundary it is given by the light front sphere. This point-sphere correspondence is an interesting  non-local correspondence that appears in our the holographic description of radiation. 
In this section we describe this point-sphere correspondence more precisely. 

 \FIGURE{\epsfig{file=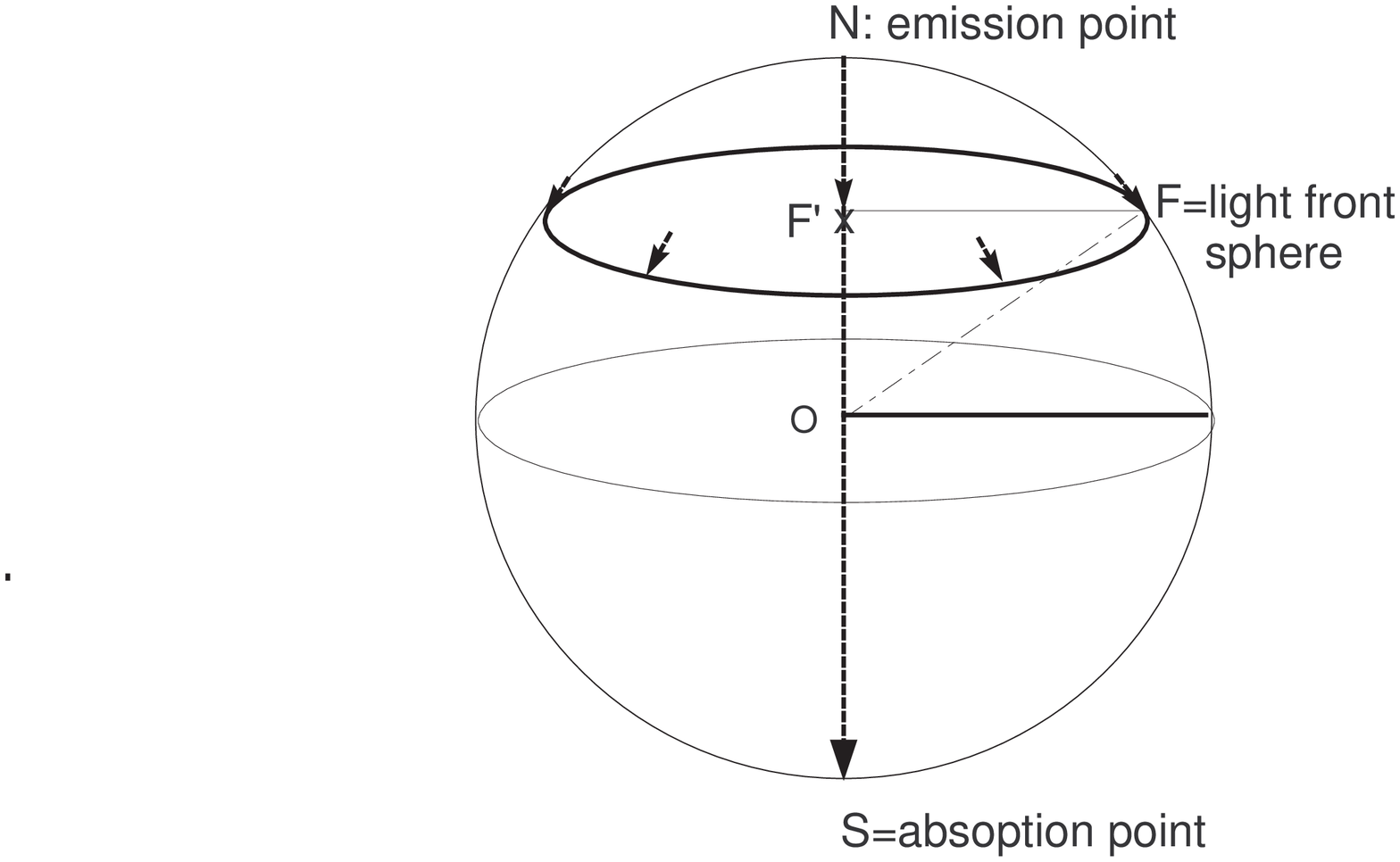,width=10cm}
         \caption[fig]{Light propagation in the bulk and boundary.  Each point of the light strip in AdS correspond to a light  front $S^2$. When the 'idea' of light arrive at F', its shadow in our world arrive at F as a spherical light front,
        such that $\psi+\theta={\pi\over 2}$ holds.
           \label{lightNS}}}

We evaluate the flight-time for IL from 
the north pole at $r=\infty$  to a position $F'$ at $r=r_0$ on the z-axis. Let the metric be 
\be
ds^2=-(1+r^2/R^2)dt^2+\frac{dr^2}{1+r^2/R^2}+r^2d\Omega^2.
\ee 
Then from the null condition of the light propagation at the bulk,
 \be
\Delta t=-\int_\infty^{r_0}\frac{dr}{1+r^2/R^2}.\ee By a change of
variable $r=R\tan\psi$, we get $ \Delta t
=R(\frac{\pi}{2}-\psi_0),$ with $\psi_0=\tan^{-1}r_0/R$. In order
to clarify the meaning of this result, we rewrite the metric in
terms of $\psi$: \be ds^2=\sec^2\psi\(-dt^2+R^2(d\psi^2+\sin^2\psi
d\Omega_2^2)\).\ee By looking with fixed azimuthal angle, we can
set $d\Omega^2=d\theta^2$.  Then the light propagation in the
bulk, when the deviation in the boundary is for infinitesimal
time, is characterized by $ \theta=0$ \cite{mik}. This together
with the null condition  leads us to \be dt=\pm Rd\psi.\ee The
boundary conditions are $\psi=\pi/2$ at $t=0$ and ${d\psi\over
dt}<0$. Therefore \be t=R({\pi\over2}-\psi).\label{bulk}\ee

On the other hand, if we look from the boundary point of view, the metric is locally Lorentz invariant since the limit $r\to \infty$
should be taken: \be ds_B^2=const.( -dt^2+dx^2),\ee with
$dx:=Rd\theta$. Since the light propagation in the vacuum must be
Lorentz invariant, its trajectory has to be the null line on the
boundary as well. This gives \be t=R\theta .\label{bdy} \ee

 By identifying the light propagation  in the
bulk and that in the boundary,   we get the relation
  \be \psi(t)=\frac{\pi}{2}-\theta(t) , \;\;{ i.e,
}\;\;\; {r(t)}={R}\cot\theta(t). \label{holodyn}\ee 
By forgetting the dynamics of light propagatin, namely the time dependence, we get the  precise form of  point-sphere correspondence,
 \be \theta=\frac{\pi}{2}-\psi   . \label{holo}\ee 
Namely, this is the   holographic
correspondence   between $\psi$, the position of  IL in the bulk and $\theta$, the light front at the boundary. 
Morally speaking, the 'idea of light' in
the bulk, draw a shadow to the boundary in the form of a spherical light-front $S^2$ in the specific way given by (\ref{holo}). We illustrated this idea in Fig.~\ref{lightNS}. 

\subsection{Radiation by a single quark in the flat boundary}
To get a physical intuition of what the point-sphere correspondence, we work out the same analysis for the flat
boundary case. 
 The flat boundary case, the geometry describes the
radiation by a single quark since its anti-quark counterpart sits
at infinity. The radiation is emitted from the origin and
propagates to infinity, a generic situation of radiation by a
charged particle, say a high-energy jet. This case is important as it carries to theories defined in flat space.

In the Poincare coordinate ($r=R^2/z$), the metric  and the action
are given by \be ds^2=\alpha'R^2\[-dt^2+ dx^2+dz^2\]/z^2.
\label{r3bdym}\ee  The light strip in the bulk in the limit of
zero width follows the null line $ -dt^2+dz^2=0,$  with solution
$z=\pm t$. The light propagation in the boundary is given by
$dz=0$ and $z\to 0$: $ -dt^2+d\vec{x}^2=0,$ with the solution $
|\vec{x}|=t.$ With this in mind, we can identify the
correspondence of light propagation in the bulk and in the
boundary: \be |\vec{x}(t)|=z(t).\ee See Fig. \ref{lightFL}. From
$z=R^2/U$, we see that     
 the  point-sphere correspondence of radiation  is  reduced to the well known UV/IR relation of Susskind and Witten \cite{susskind}.  

\section{Geometry of screening length}

Now we turn to the primary goal of this paper, which is the gluon screening in the hot medium. The finite temperature Yang-Mill theory in Ads/CFT this is realized  by putting  a black hole in the AdS bulk\cite{witten2}. The relevant metric is given
by~\cite{witten2,sonn,rey2}, \be ds^2=-f(r)dt^2+{dr^2\over
f(r)}+r^2d\Omega^2,\ee with $f(r)=1+r^2/R^2-w/r^2$  and
$w=8GM/3\pi$. Following the steps above, the light propagation in
the bulk follows along the axis $\theta=0$ and the flight time is
\be \Delta
t=-\int_\infty^{r_0}\frac{dr}{1+r^2/R^2-w/r^2}.\label{bhbulk}\ee
Let $r_+$ be the positive solution of $f(r_+)=0$.
 Notice that the
flight time diverge as $r_0$ moves to the horizon at $r=r_+$.

On the boundary, however, the gluon propagation in the strongly interacting medium  at finite temperature is not  something that allows any easy treatment. So the hope is if we know the bulk-boundary correspondence a-priory, then we can read off the light front of gluon from the position of IL by use of the correspondence. 

Although the precise form of point-sphere correspondence in the presence of the black hole is not clear, it is quite obvious that the the IL can not go through the black hole, no matter how small is the black hole, because the IL has to pass the center.  
As is clear form the eq. (\ref{bhbulk}), it can never reach the horizon in finite time, namely it stalls near the horizon. If the IL should stall, so should do its shadow at the boundary.
Therefore there is a finite distance beyond which gluon can not propagate further. This is the mechanism for the complete shielding of the gluon in the  strongly interacting hot medium. 

To describe this idea more quantitatively, we make an assumption that the holographic correspondence is not sensitive to the deformation of the metric in the bulk as far as the asymptotic form remains the as AdS space. In other words, we assume that 
the point-sphere correspondence given by eq.(\ref{holo})
is valid in the presence of the black hole. 
This ansatz is motivated from the fact that point-sphere correspondence is a correspondence between position at the bulk and the size at the boundary. Position in the bulk can be defined without referring to the metric  and there is no metric dependence in the boundary hence size at the boundary has a universal meaning without refereing to the metric in the bulk either. 
Under this ansatz,  the bulk-boundary relationship 
\be
\psi+\theta=\pi/2\ee
in the presence of black hole remains the same as the pure AdS case although the dynamics of 'how IL move in the bulk' is different from the pure AdS case. 
We call this ansatz as the "metric independent point-sphere "(MIPS) correspondence.

If one accept MIPS correspondence, the  description of the boundary propagation is very simple.  
For ads black hole background, it takes
infinite time for the IL to arrive at the horizon $r=r_+$.
Correspondingly, the light front in the boundary arrives at the
maximal distance from its emission point N: \be
L_c=R\theta_c=R\cot^{-1}(r_+/R)\ee beyond which light cannot
propagate. This is the desired stopping/screening distance of radiation. Just as the
asymptotic observer never sees the entrance of light in the black
hole, the gluon does not propagate further than the
stopping/screening length. The latter is a time-like screening
length which is different from the space-like screening length
discussed in~\cite{sonn,rey2}. While the latter sets the range of the force in the medium, the former sets the range of the mean-free 
path for travelling light in a medium. Clearly the
stalling of light in a strongly coupled medium simply
means that the energy carried by the wave dissipates through Ohm's
law in the medium, and the wave ceases to exist as such.

Now, with this understanding, we can express the bulk relation
between $dt$ and $dr$ in terms of  a boundary relation between
 $dt$ and $dx$ with $x=R\theta$: \be dt=n(\theta) Rd\theta =n(x) dx, \ee
  with
\be
n(\theta)=\frac{1}{1-\frac{w}{R^2}{\sin^2\theta}{\tan^2\theta}}.
\ee Notice that since $w$ is proportional to the black hole mass,
the refraction will disappear in the zero mass limit.
\footnote{One subtle point of the black hole physics is that the
AdS is not the zero temperature limit of the AdS black hole, but
the zero mass limit. In the $M\to 0$ limit, $T\to 1/(2\pi r+)$ and
$r_+ \to \sqrt{w/R^2}$. In the flat boundary case, the zero
temperature limit is the zero horizon limit, since $T\sim r_T$. }

\FIGURE{\epsfig{file=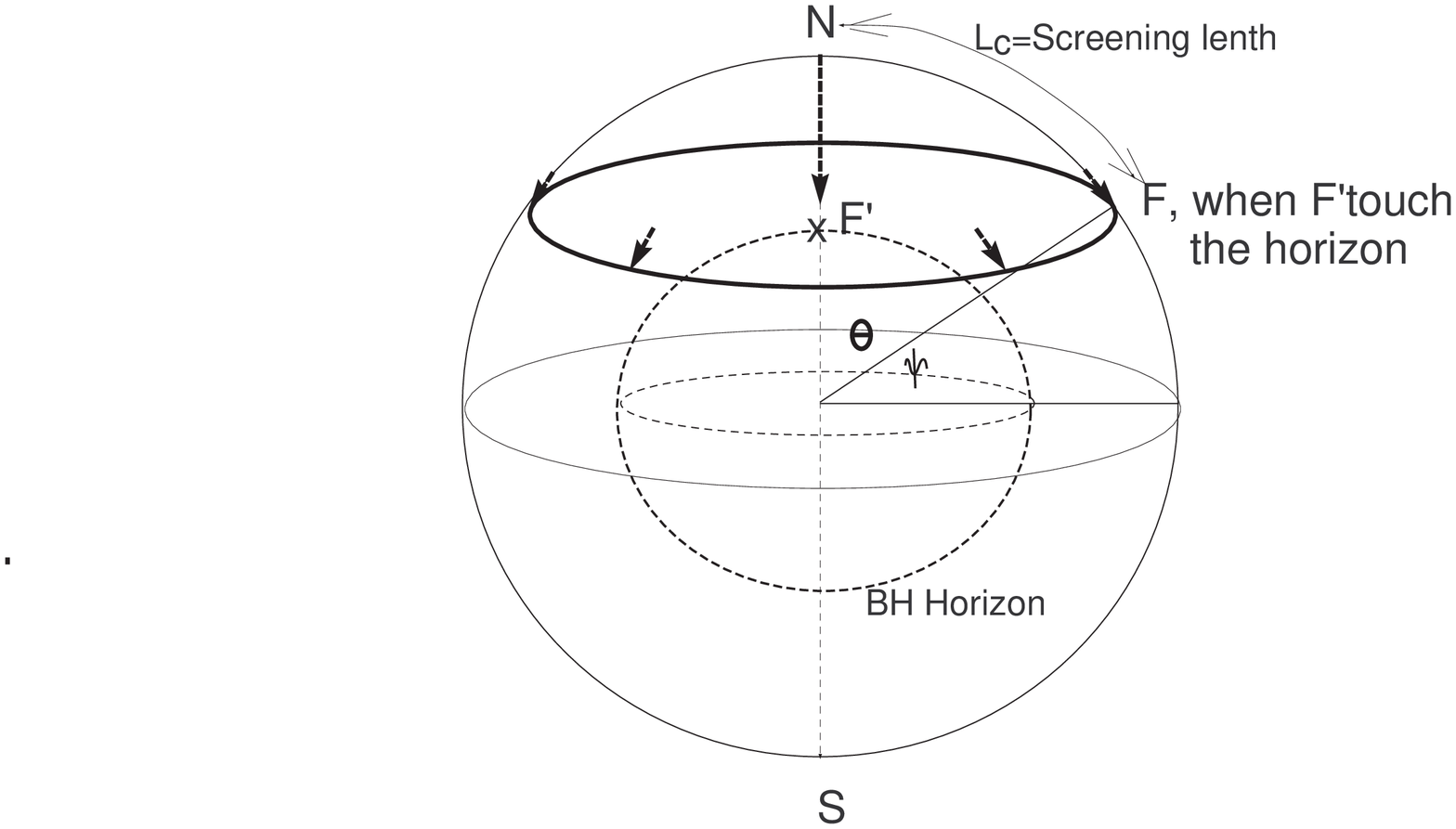,width=10cm}
        \caption[fig]{Light propagation in the presence of black hole.
          When 'Idea' of light arrive at the horizon of the black hole,
          the light at the boundary approach to its screening
          length $L_c$.
              \label{lightBH}}}
Notice that since the velocity of the light (gluon) propagation in
the boundary is \be v_g=\frac{dx}{dt}=\frac{1}{n(x)},\ee we may
interpret $n(x)$  as the `refraction' coefficient characterizing
the medium. The dependence of the refraction coefficient on
$\theta$ does not imply anisotropy of the medium. It just reflects
the dependence of the light propagation on the distance from the
emission point. The velocity of light ($1/n$) slows down to zero
as it approaches  the screening distance $L_c$. See figure
\ref{lightBH}.

\subsection{Flat boundary case}

Now we consider the case with black hole background, corresponding
to finite temperature Yang-Mill theory. The metric is given by \be
ds^2=\frac{\alpha'R^2}{z^2}\[-(1-{z}^4/{z_T}^4)dt^2+
dx^2+\frac{dz^2}{ (1-{z}^4/{z_T}^4)}\].\label{r3bdyBHm}\ee The
radiation strip  in the bulk follows the null curve in the narrow
limit  \be -(1-{z}^4/{z_T}^4)dt^2 +\frac{dz^2}{
(1-{z}^4/{z_T}^4)}=0,\ee with solution \be
t=\frac{z_T}{2}\[\tanh^{-1}\frac{z}{z_T}+\tan^{-1}\frac{z}{z_T}
\].\ee
Near $z\sim 0$, $t\sim z$ and when $t\to \infty$, $z$ approaches
$z_T$.

\FIGURE{\epsfig{file=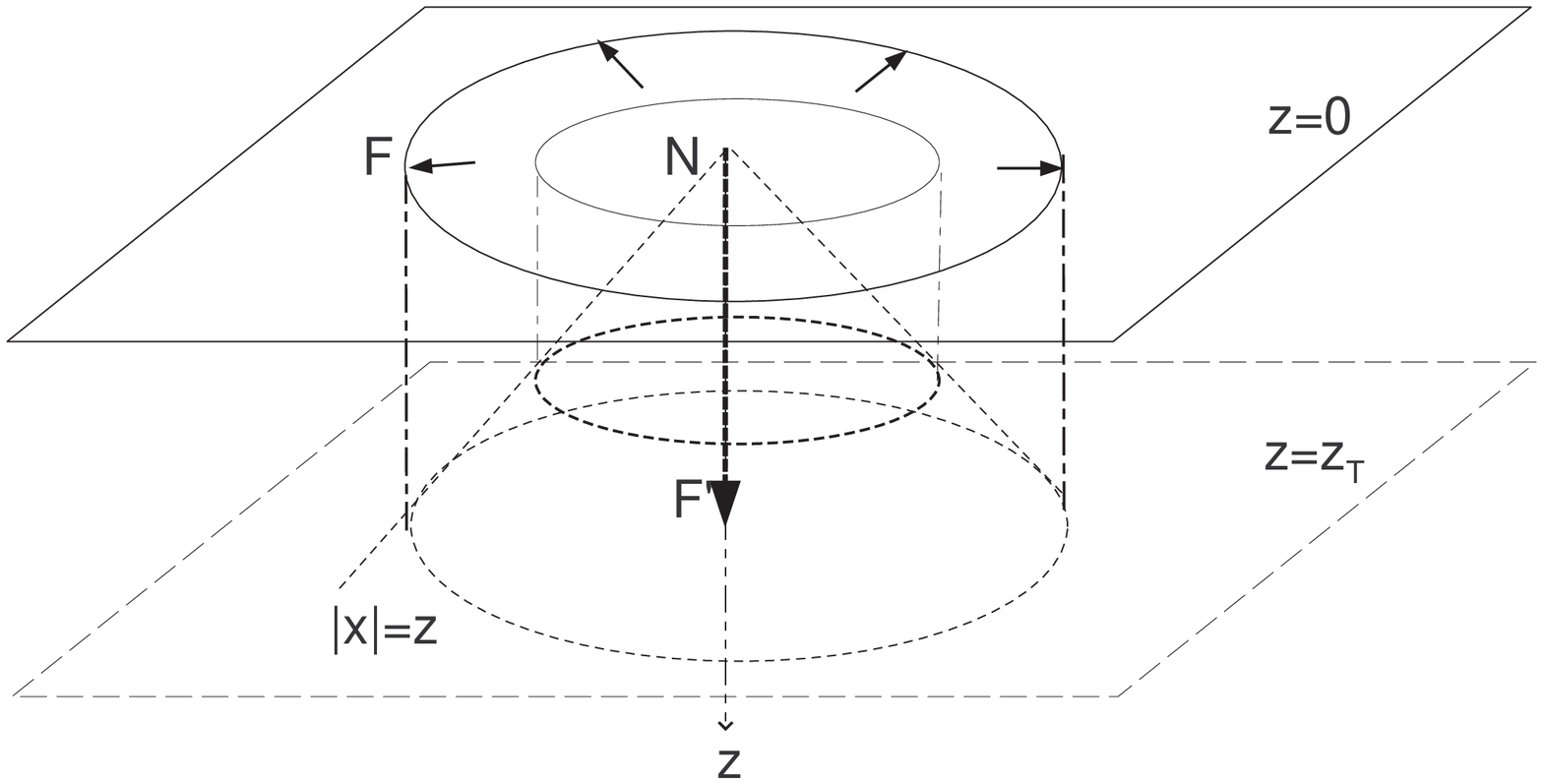,width=10cm}
       \caption[fig]{Propagation of radiation emitted by a single
      quark at the North pole N in the flat boundary.
     In a finite temperature medium, the bulk geometry has a
    horizon at $z=z_T$.
   \label{lightFL}}}

The light propagation in the boundary is given by $dz=0$ and we
have to introduce the 'index of the refraction' $n$ through
$dt-n(x)\cdot d|\vec{x}|=0.$  With the identification $
|\vec{x}|=z,$ we can identify the correspondence of light
propagation in the boundary as: \be
\frac{d|\vec{x}(t)|}{dt}=\frac{1}{n(x)}= 1-\frac{|
\vec{x}|^4}{z_T^4}=1-(\pi T |\vec{x}|)^4.\ee See
Fig.\ref{lightFL}. Just as the the falling of IL is limited by the
horizon at $z=z_T$, the propagation of light in the finite
temperature medium is limited by \be |\vec{x}|=z_T=\frac{1}{\pi
T}, \label{ls}\ee where $T$ is the temperature. We call this: the
screening distance of radiation $L_c^{rad}$. \footnote{The idea of
duality between thermalization in boundary  and free fall in the
bulk is stated first in \cite{bala} in the context of radius-scale
duality and further discussed in \cite{dani}.}

If we consider our choice MIPS-correspondence literally, we can compare the screening distance of radiation with that of static forcec. It turns out that MIPS-correspondence gives the  former  larger than  the latter, which was introduced in \cite{sonn,rey2}. The latter is the distance over
which energy of two parallel strings are less than that of the
U-shape. From the expression of separation distance \be L=2z_T
(1-\epsilon)^{1/4}\sqrt{\epsilon}\int_1^\infty
\frac{dy}{\sqrt{(y^4-1)(y4-1+\epsilon )}},\ee with
$\epsilon=1-U_T^4/U_0^4, \;z=R^2/U$. It can be worked out
numerically that the critical value of $\epsilon$ is
$\epsilon_c=0.81445$ and the screening distance of a static force
is \be L_c^{static}=2z_T\cdot 0.37705\bra L_c^{rad}.\ee
 That is, the radiation can reach farther than the static
field by about $33 \%$ in the strongly interacting finite
temperature medium.  In \cite{dani},  {\it static} screened
  wave amplitude was derived and the static screening length
calculated there is $L_S= 0.29z_T$, which is even smaller than
$L_c^{static}$ and hence much less than the radiation screening
length given in eq.(\ref{ls}). One may regard this feature of nonlinear radiation as the reminiscent of $U(1)$ electrodynamics in free space.

\section{Discussion}

We have considered the holographic correspondence of the
propagation of non-Abelian waves in vacuum and finite temperature
assuming the holographic principle.  We pointed out that the light propagation in the bulk is described by a point like object passing through the center of the AdS space, which, in the presence of the black hole, unavoidably stalls at the horizon. 
We have suggested that this is the mechanism why the radiation stalls at a distance of $1/\pi T$ in the finite temperature
medium irrespective of how much energy it carries. The latter is just dissipated in the form of heat throughout the medium (Ohm law).  
To arrive at this conclusion we used the metric independent point-sphere correspondence. However, the gluon screening mechanism suggested in this paper is independent of precise form of the point-sphere bulk-boundary relation. 

We also have suggested that these observations in ${\cal N}=4$ may carry to QCD near its deconfining temperature, as it is likely
to be in a strongly coupled quark-gluon liquid. In
particular, the stalling of the colored non-Abelian waves in the
medium at a distance of order $1/\pi T_c\approx 1/3$ fm whatever
the radiation energy, would imply that the quark-gluon liquid is
very 'opaque'. High energy jets at RHIC would not make it beyond
$1/3$ fm. This implies that the quark-gluon liquid has a very
large color conductivity resulting into a skin depth of about
$1/3$ fm. The 'opaque' character of the quark-gluon liquid
whatever the energy, yields a first principle explanation to
the large suppression of high energy jets at RHIC. Indeed, the
suppression is found to be 'oblivious' to the energy of the jet.

Finally, we note that our analysis for the wave-propagation was
carried in the geometrical limit ignoring effects related to
diffraction and dispersion. This treatment is justified  in the Maldacena limit of infinitly strong t'
Hooft coupling. It would be worth investigating the effects of
diffraction on our results by relaxing this limit. In fact, It
will be interesting if our result can be derived and extended
using the method of ref.\cite{dani}. We will report the result in
a forthcoming paper.

\bigskip

\acknowledgments

SJS thanks Andrei Mikhailov for explaining his work. IZ thanks
Hanyang University for hospitality during the initial stage of
this work. We both express our thanks to Mannque Rho for his
support and interest. The work of SJS was supported   by the KOSEF Grant R01-2004-000-10520-0 and the work of IZ is supported in US-DOE grants DE-FG02-88ER40388 and DE-FG03-97ER4014.


\end{document}